\documentstyle[epsfig,graphicx,12pt,epsf]{article}

\newcommand{\be}{\begin{equation}}
\newcommand{\ee}{\end{equation}}
\newcommand{\ba}{\begin{eqnarray}}
\newcommand{\ea}{\end{eqnarray}}

\begin{document}

\begin{center}
{\Large{\bf Pionic Atom Formation in Halo Nuclei by
(d,$^3$He) Reactions}}

\vspace{0.3cm}

\end{center}

\vspace{1cm}

\begin{center}
{\large{M. Fujita, S. Hirenzaki and K. Kume }}
\end{center}

\begin{center}
{\small{ \it Department of Physics, Nara Women's University, Nara
630-8506, Japan}}

\end{center}

\vspace{1cm}

\begin{abstract}

We investigate the formation cross sections of the deeply bound pionic
atoms on halo nuclei theoretically.  We consider one-neutron pick-up
(d,$^3$He) reaction for the pionic atom formation and find that the
formation rate is significantly increased by the existence of the halo
neutron with long tail of the wave function.  This enhancement is expected
to increase the experimental feasibility for the formation of the deeply
bound pionic states on $\beta$-unstable nuclei.

\end{abstract}

\section{Introduction}

 Mesic atoms are the useful laboratories for
 studying the meson-baryon interactions, the meson
  properties in nuclear medium and the nuclear properties.
 Among others, pionic atoms have been long utilized
for such studies \cite{ericson88}.
The binding energies and widths
of the pionic bound states are expected to provide us with the unique
 information of the pion-nuclear interaction \cite{laat91}
 and of the nuclear density distributions \cite{hirenzaki87,toki90}.

The standard method to produce pionic atoms
starts with
injecting slow negative pions into matter.  The pions
will be stopped and  trapped in outermost orbits of atoms,
 then lose
 energy by emitting Auger electrons and x-rays to cascade down to
deeper atomic states.
 However, the x-ray cascade ceases at the so called
 'last orbital', where the pions are absorbed by the nucleus without
going to deeper atomic states.
This prevents us from studying
 the deep pionic states for heavy nuclei that
 have large overlap with nuclear density \cite{toki89}.

 To overcome this difficulty,
 the recoilless mesic atom production by
(d,$^3$He) reaction has been proposed \cite{hirenzaki91}.
This method tries to directly create deeply bound pions
by the nuclear reaction without cascade down of the pion
 from the outer most orbits.
This direct method was proved to be
effective and  powerful for the study of
far deeply bound pionic atoms \cite{yamazaki96,gillitzer00}.
We can now determine the binding energies and widths of
the deepest bound states such as 1s and 2p atomic states
 from the energy spectra of the emitted $^3$He.
 We remark that the (d,$^3$He) reaction is so powerful that
it may also be applied to the creation of
neutral meson-nucleus bound states  such as  the $\eta$ and
$\omega$ mesic nuclei \cite{hayano99,hirenzaki02}.

On the other hand, the physics of the unstable nuclei is one of the most
interesting subject in modern nuclear science.
Since the successful use of a new experimental tool,
namely ``beams of unstable nuclei'' \cite{tanihata85,tanihata89},
the properties of nuclei far from the stability line have been studied
extensively in many laboratories where secondary beams of unstable nuclei
are available.
One of the exciting findings was the neutron halo structure around the
$^9$Li core in $^{11}$Li \cite{tanihata85}.
Since then, the structure and reactions of $\beta$-unstable nuclei
have been studied extensively both theoretically and experimentally
\cite{tanihata96,horiuchi01}.

Since
the repulsive S-wave pion-nucleus interaction causes the formation of
the pionic halo around the core nucleus \cite{toki89},
the binding energies and widths of deeply
bound pionic atoms are expected to be very sensitive to the structure of the
nuclear surface,
in particular, to the neutron skin \cite{toki90}.
Hence,
we can expect to obtain new information on neutron/proton
distribution of the unstable nuclei by observing the deeply bound
pionic atoms.
The determination of  the neutron distributions for unstable nuclei
is one of the most important subjects in the $\beta$-unstable
nuclear physics \cite{tanihata96}.
In addition the (d,$^3$He) reaction with pionic atom formation
proceeds through the single-neutron pick up and
are known to be very sensitive
to the single-particle properties of the neutron in nucleus \cite{umemoto00}.
Hence, we can also expect to obtain the new information of
the neutron single-particle states $(nlj)$ in unstable nuclei.
This is very interesting in the context of the possible change
of the magic numbers far from stability line \cite{tanihata98}.

For the formation of the deeply bound pionic atoms on unstable
nuclei, we use the inverse kinematics technique, in which
a light target such as deuteron is bombarded by radioactive ion (RI) beam of
intermediate energy and the recoiling light ejectile $^3$He is detected
after forming pionic atoms in the heavy projectile
\cite{yamazaki90}.
We evaluate the formation cross sections for inverse kinematics reactions
d(RI,$^3$He)X in order to examine the experimental feasibility
\cite{umemoto01}.
Recently, the existence of the halo like structure of neutron
distribution is predicted theoretically in medium heavy mass
nuclei like Zr isotopes \cite{meng98}, which was not considered
in our previous work \cite{umemoto01}. This is very
interesting since the existence of the long tail of the neutron
wave function may enhance the formation cross section of the
deeply bound pionic atoms in unstable nuclei and increase
much the experimental feasibility in
very unstable region of nuclear chart using the next generation
of accelerators such as RIKEN RIB factory.

 In this paper
we study the formation cross sections of the deeply bound pionic
states for the medium and heavy halo nuclei.
We simulate the halo-like structure by using the solutions of the
Schr\"odinger equation with an one-body neutron potential for various
potential depths, and we investigate the effects of the long-tail of the
neutron wave functions to the (d,$^3$He) spectra.
In section 2, we describe our theoretical formalism. We show the
numerical results in section 3. Section 4 is devoted to the summary.

\section{Formalism}

In this section, we describe our theoretical formalism to investigate the
pionic atom formation spectra on halo nuclei.  As we will describe below, we
introduce a single free parameter $F$, a reduction factor for neutron potential
depth, and simulate the halo structure by changing the parameter $F$.
Nuclear properties, such as neutron wave functions, density distribution and
one-neutron separation energies, are obtained consistently for various
$F$ values.  Pionic atom structure and formation spectra are also
calculated consistently for several $F$ values.
The factor $F$ is the only free parameter introduced in the present work.

In order to simulate the halo wave functions for neutron,
we solve the one-body Schr\"odinger equation for all occupied
single-particle states for both protons and neutrons
using the potential in ref. \cite{speth77}. The one-body potential for
protons and neutrons are given by,

\begin{eqnarray}
{V\left(r\right)}
&=& V_{0}{{1}\over{1+{\rm exp}\left(\left(r-R_{0}\right)/\alpha_{0}\right)}}
\nonumber\\
& &-{{\hbar}^2\over{2M^2c^2}}\lambda_{ls}\left(
\mbox{\boldmath$\ell$}\cdot\bf{s}\right){{1}\over{r}}{{d}\over{dr}}
\left({V_0}\over{1+{\rm
exp}\left(\left(r-R_{ls}\right)/\alpha_{ls}\right)}\right)+V_c(r),
\label{eqn:speth}
\end{eqnarray}

\noindent
where $V_{c}$ is the Coulomb potential of a uniform charge distribution with the
radius $R_{c}$.
We adopt the potential parameters which were determined so as to fit the experimental
single particle energies for Pb region in ref. \cite{speth77} and these are listed in Table \ref{table1}.

Here, we introduce a free parameter $F$ which is a reduction factor for
neutron potential
depth to simulate the halo structure as;

\begin{equation}
V_0 \rightarrow F \times V_0 .
\label{eqn:factor}
\end{equation}

\noindent
This factor is introduced only for the neutron potential and
the potential depth for proton is not altered for all numerical calculations in the present work.

Using the proton and neutron radial wave functions
$R_{nlj}^{(p/n)}(r)$, we calculate the
proton and neutron density distributions by summing up the absolute square of
the wave functions for all occupied states $(nlj)$ for proton and
neutron, respectively,

\begin{equation}
\rho_{(p/n)}\left(r\right)
={{1}\over{4\pi}}\sum_{nlj}{\left(2j+1\right)|R_{nlj}^{(p/n)}\left(r\right)|
^{2}}.
\label{eqn:pndens}
\end{equation}

\noindent
We calculate the density of the target nucleus $\rho_{A}$ using Eq.(\ref{eqn:pndens}), and we evaluate the final-state nuclear density $\rho_{A-1}$, which appears in  Eq.(\ref{eqn:distortion}), by removing one neutron in the state which has the largest contribution to the (d,$^3$He) spectra from the $\rho_{A}$.

The pionic atom wave functions and eigen energies are obtained by
solving the Klein-Gordon equation with pion-nucleus optical potential,

\begin{equation}
{
\left[-\nabla^{2}+\mu^{2}+2\mu V_{opt}\left(r\right)\right]\phi_{\pi}\left(
\bf{r}\right)=\left[E-V_{Coul}\left(r\right)\right]^{2}\phi_{\pi}\left(\bf{r}\right)
},
\end{equation}

\noindent
where $\mu$ is the reduced mass of pion-nucleus system and $V_{Coul}(r)$ is
the Coulomb
potential with a finite nuclear charge distribution $\rho_{p}\left(r\right)$
given by,

\begin{equation}
{
V_{Coul}\left(r\right)=-e^{2}\int{\rho_{p}\left(r'\right)\over|\bf{r}-\bf{r'
}|}d^{3}r'
}.
\end{equation}

\noindent
We use the proton and neutron density distributions $\rho_{p}$ and
$\rho_{n}$ of target nucleus given in Eq. (\ref{eqn:pndens}) for all
densities in the Klein-Gordon equation.
Hence, all nuclear densities and the pion wave function
depend on the parameter $F$.

For the pion-nucleus optical potential, we adopt the conventional Ericson-Ericson form \cite{ericson66};

\begin{equation}
{
2\mu V_{opt}\left(r\right)=-4\pi\left[b\left(r\right)+\epsilon_{2}B_{0}
\rho^{2}\left(r\right)\right]+4\pi\nabla\cdot\left[c\left(r\right)+
\epsilon_{2}^{-1}C_{0}\rho^{2}\left(r\right)\right]L\left(r\right)\nabla
},
\end{equation}

\noindent
with

\begin{equation}
{
b\left(r\right)=\epsilon_{1}\left[b_{0}\rho\left(r\right)+
b_{1}\left[\rho_{n}\left(r\right)-\rho_{p}\left(r\right)\right]\right]
} ,
\end{equation}
\begin{equation}
{
c\left(r\right)=\epsilon_{1}^{-1}\left[c_{0}\rho\left(r\right)+
c_{1}\left[\rho_{n}\left(r\right)-\rho_{p}\left(r\right)\right]\right]
} ,
\end{equation}
\begin{equation}
{
L\left(r\right)=\left[1+{4\over3}\pi\lambda\left[c\left(r\right)+
\epsilon_{2}^{-1}C_{0}\rho^{2}\left(r\right)\right]\right]^{-1}
}.
\end{equation}

\noindent
We adopt a parameter
set determined by Seki and Masutani in ref. \cite{seki83} as the
standard one and these are listed in Table \ref{table2}.

To evaluate the formation rate of the pionic atoms in the (d,$^3$He)
reactions, we use the effective number approach \cite{umemoto00} which is
known to predict the experimental cross sections reasonably well
\cite{yamazaki96}.
The (d,$^3$He) spectrum is given as;

\begin{equation}
\left( \frac{d^2\sigma}{d\Omega_{\rm{He}} dE_{\rm{He}}} \right)_{dA \rightarrow ^3
\rm{He} (A-1) \pi}= \sum_{[l_\pi \otimes j_n^{-1}]}
\left( \frac{d \sigma}{d \Omega} \right)_{dn \rightarrow ^3\rm{He} \pi}
N_{ {\rm eff}} \frac{\Gamma _{\pi a}}{2 \pi}
\frac{1}{\Delta E ^2 + \Gamma _{\pi a}^2/4},
\label{eqn:cross}
\end{equation}

\noindent
with

\begin{equation}
\Delta E = Q + m_{\pi} - BE_{\pi} + S_n - [M_d + M_n - M_{\rm{He}}] ,
\label{eqn:deltae}
\end{equation}

\noindent
where $BE_{\pi}$ and $\Gamma _{\pi a}$ are the binding energy
and the width of pionic atom, and $S_n$ is the neutron separation energy. These quantities
also depend on the factor $F$ introduced in Eq.(\ref{eqn:factor}).

The effective number $N_{ {\rm
eff}}$ in Eq.(\ref{eqn:cross}) is defined with the neutron and the pion wave functions
$\psi_n\left(\bf{r}\right)$
and $\phi_\pi\left(\bf{r}\right)$ as,

\begin{equation}
{
 N_{{\rm eff}}=\sum_{JM} \biggm|
\int\chi^*_f \left(\bf{r}\right)
\left[\phi^*_\pi\left(\bf{r}\right) \otimes
\psi_n\left(\bf{r}\right)\right]_{JM}
\chi_i \left(\bf{r}\right)\  d^{3}\bf{r}
\biggm|^2
} ,
\end{equation}

\noindent
where the deuteron and $^3$He wave functions are expressed as $\chi_i$ and
$\chi_f$, and are calculated in Eikonal approximation as;

\begin{equation}
{
\chi^*_f \left(\bf{r}\right)\chi_i \left(\bf{r}\right)= \exp\left(i\bf{q}
\cdot \bf{r}
\right)D\left(z,\bf{b}\right)
}.
\label{eqn:eikonal}
\end{equation}

\noindent
The distortion factor
$D\left(z,\bf{b}\right)$ appeared in Eq. (\ref{eqn:eikonal}) is defined
as;

\begin{equation}
{
D\left(z,\bf{b}\right)\
=\ \exp\left[-\frac{1}{2} \sigma_{dN}
\int_{-\infty}^z dz'\ \rho_A\left(z',\bf{b}\right)
-\frac{1}{2}\sigma_{hN}
\int_z^\infty dz'\ \rho_{A-1}\left(z',\bf{b}\right)\right]
},
\label{eqn:distortion}
\end{equation}

\noindent
where the nuclear densities are given in Eq. (\ref{eqn:pndens}).

We like to mention here again that we have only a single free parameter $F$ in
this model and all physical quantities are obtained in a consistent manner
for each $F$ value
to simulate the effects of the halo structure on the pionic atom formation
(d,$^3$He) spectra.

\hspace{-1cm}
\begin{table}[h]
\begin{tabular}{cccccccc}
\cline{1-8}
  & $V_{0}$(MeV) & $R_{0}$(fm) & $\alpha_0$(fm) & $\lambda_{ls}$ & $R_{ls}$(fm) &
$\alpha_{ls}$ (fm)& $R_{c}$(fm) \\  \cline{1-8}
protons & -60.94 & 7.52 & 0.79 & 22.33 & 7.28 & 0.59 & 6.70\\
neutrons & -49.40 & 7.05 & 0.66 & 21.47 & 6.78 & 0.24 &  \\ \cline{1-8}
\end{tabular}
\caption{Parameters for proton and neutron one-body potential obtained in
ref. \cite{speth77} for Pb region.}
\label{table1}
\end{table}

\begin{table}[h]
\begin{center}
\begin{tabular}{ll}
\cline{1-2}
$b_0$= - 0.0283$m_{\pi}^{-1}$ & $b_{1}$= - 0.12$m_{\pi}^{-1}$\\
$c_0$=0.223$m_{\pi}^{-3}$ & $c_{1}$=0.25$m_{\pi}^{-3}$\\
$B_0$=0.042 $i$ $m_{\pi}^{-4}$ & $C_{0}$=0.10 $i$ $m_{\pi}^{-6}$\\
$\lambda$=1.0 & \\
\cline{1-2}
\end{tabular}
\caption{Pion-nucleus potential parameters obtained in ref.
\cite{seki83}. }
\label{table2}
\end{center}
\end{table}

\section{Numerical results}

In this section we show the numerical results for the pionic atom formation spectra in (d,$^3$He) reaction by varying the parameter $F$ to
simulate the halo-like structure.  First, we
consider the pionic atoms on $^{208}$Pb.  The calculated neutron binding
energies are listed in Table \ref{Pb-b.e.}.
Obviously, the binding energies are smaller for smaller $F$ values.
The 1i$_{13/2}$ state does not bound for the values $F$ smaller than 0.765.
Thus, we vary the $F$ between 1 and 0.765.
These binding energies are used as the
neutron separation energies $S_n$ appeared in Eq.(\ref{eqn:deltae}).

\begin{figure}[h]
\begin{center}
\includegraphics[height=8cm]{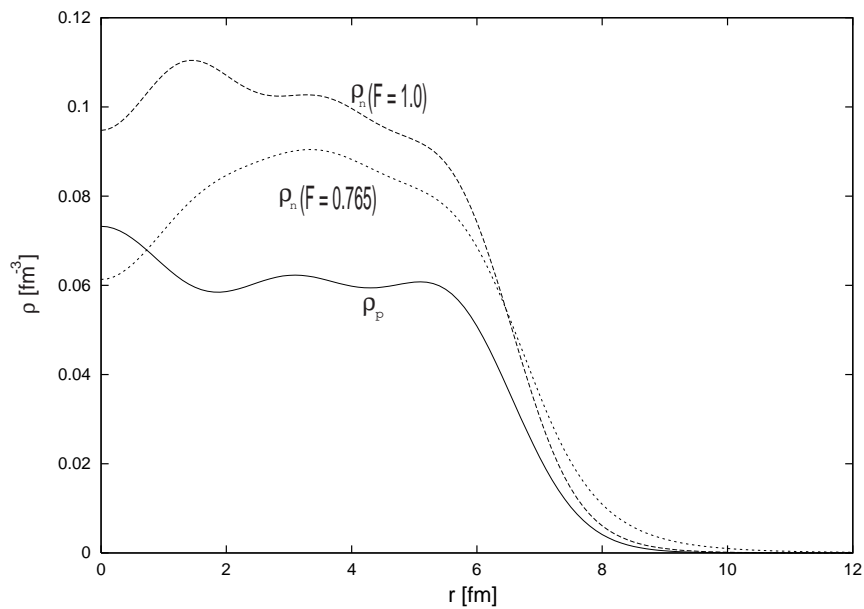}
 \caption{Neutron density distributions $\rho_{n}$ are plotted as a function of the radial coordinate $r$ for $F$=1.0 and 0.765. Proton
density  $\rho_{p}$ is also plotted as the solid line. }
\label{fig:dens}
\end{center}
\end{figure}

We show the nuclear density distributions in Fig. \ref{fig:dens} for
the values $F$ = 1.0 and 0.765.  We can see that the neutron density
distribution is affected significantly by the choice of $F$ value and has longer
tail for $F$=0.765.  The nuclear
densities shown in Fig. \ref{fig:dens} are used to calculate the pionic
wave functions.

\begin{figure}[h]
\begin{center}
\includegraphics[height=8cm]{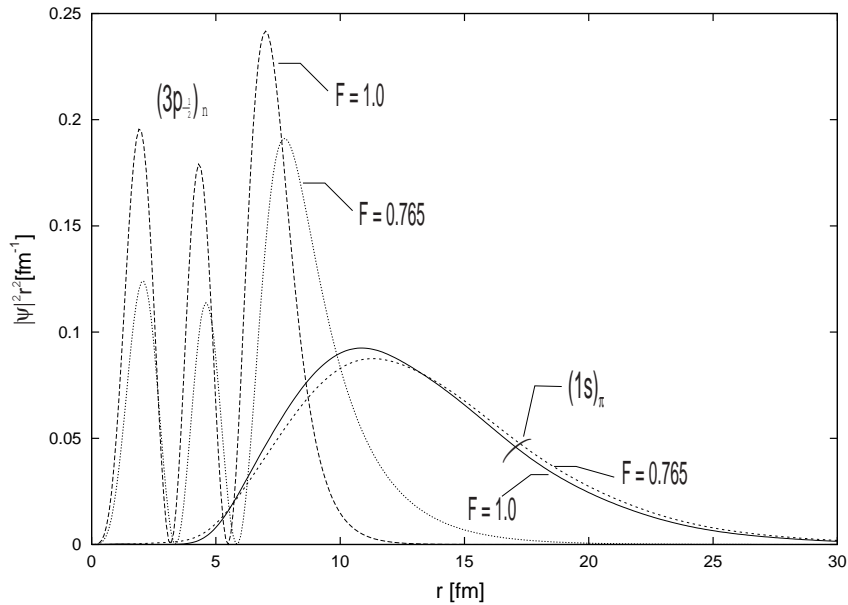}
 \caption{Density distributions of the neutron 3p$_{1/2}$ and the
 pionic 1s states are shown for $F$=1.0 and 0.765 for Pb.}
\label{fig:wave}
\end{center}
\end{figure}

The density distribution of the neutron 3p$_{1/2}$ state and that of the
pionic 1s state are shown in Fig. \ref{fig:wave}.
We can see that the neutron single-particle density in the valence
level is sensitive to the $F$ value and has halo-like tail for the smaller
$F$ values.  On the other hand, the density of the pionic atom depends 
on the $F$ value in rather mild manner as seen in Fig. \ref{fig:wave}.
The binding energy of the 1s pionic state is 6.93MeV for $F$=1 while 6.67MeV for $F$=0.765, respectively.

We also show the calculated root mean square radii for valence neutron
levels and show the dependence on the $F$ value in Table \ref{Pb-rms}.
We can see that the neutron states with lower
angular momenta are more sensitive to the $F$ value and develop the longer
tail.  This feature is interpreted as the consequence of the centrifugal
barrier
and was suggested in ref. \cite{tanihata95}. 
We expect that the long tail of the neutron distribution  leads to larger cross section for producing
 deeply bound pionic states
since the quasi-substitutional states are know to be largely populated
for the recoilless (d,$^3$He) reactions \cite{hirenzaki91}.

\begin{figure}[h]
\begin{center}
\includegraphics[height=8cm]{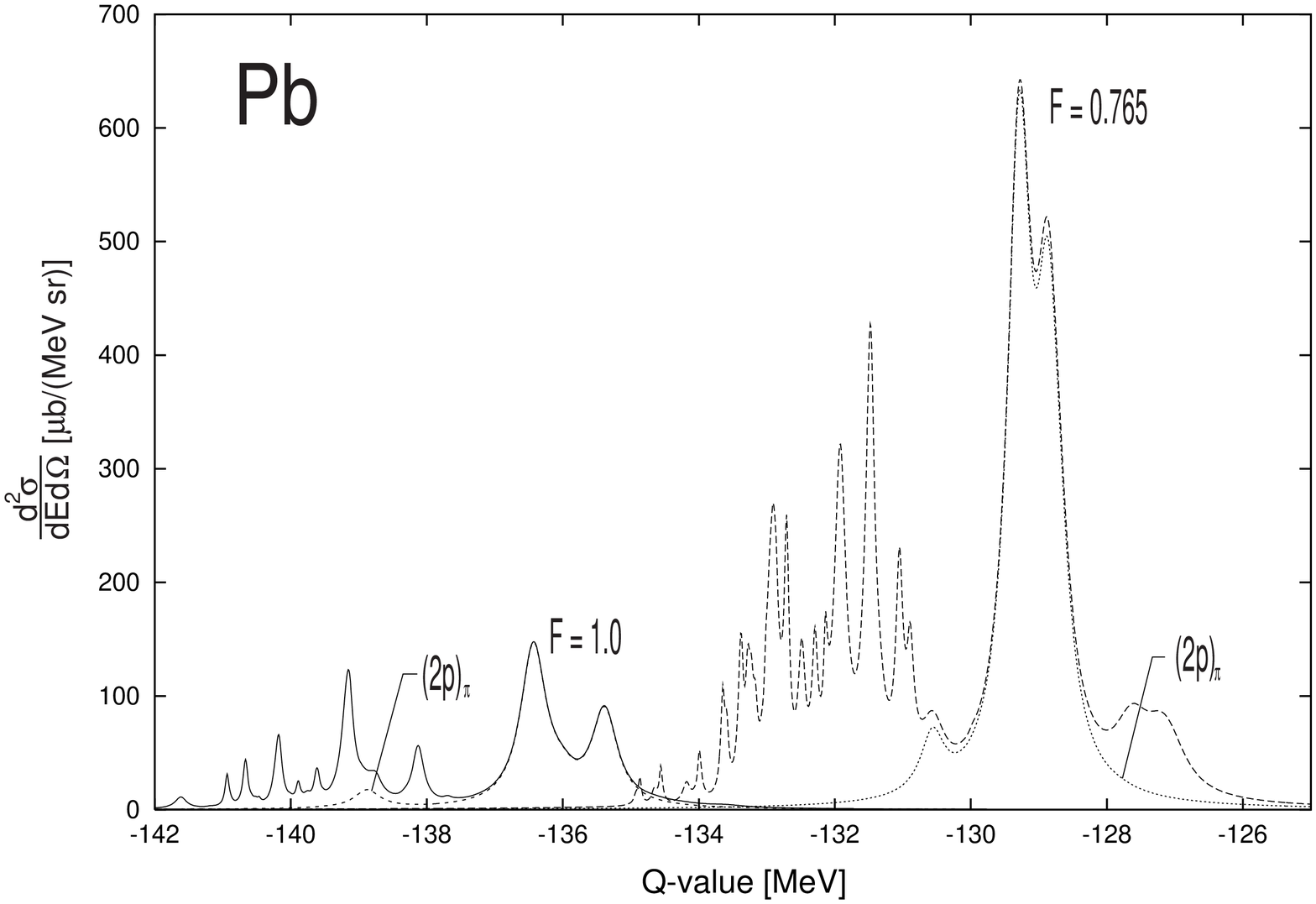}
 \caption{Formation cross sections of the bound pionic states in (d,$^3$He) reaction are plotted as a function of Q-value at the incident energy T$_d$=500 MeV on Pb.
The solid and the dashed curves are the results with $F$=1.0 and 0.765,respectively. The largest peak corresponds to the pionic 2p state, and then we show the separate contributions coming from the pionic 2p components as dotted curves. Instrumental energy resolution is assumed to be 50 keV. }
\label{fig:Pb-spect}
\end{center}
\end{figure}

To see this, we show the calculated (d,$^3$He) spectra for $^{208}$Pb target in Fig. \ref{fig:Pb-spect} as a function of Q-value. Here, we only evaluate the contributions from bound pionic states. As can be seen in the figure, we find that the formation cross section is significantly enhanced for the case of halo-like neutron distribution corresponding to $F$=0.765.

The formation cross section of [(2p$_{3/2}$)$_{n}^{-1}\otimes$(2p)$_{\pi}$] configuration, which has the largest contribution to the spectra, is around factor 3.5 larger for $F$=0.765 than that for $F$=1.0 which corresponds to no-halo case. We also find that the shape of the spectrum is considerably influenced with the change of the $F$ value.  This is due to (1) the change of the separation energy of each neutron state as shown in the Table \ref{Pb-b.e.}, and (2) the change of the momentum transfer for different Q-values. Thus, for example, we can see from Fig. \ref{fig:Pb-spect} that the energy difference between two peaks [(2p$_{3/2}$)$_{n}^{-1}\otimes$(2p)$_{\pi}$] and [(2p$_{1/2}$)$^{-1}\otimes$(2p)$_{\pi}$]  are much closer for $F$=0.765 case and there appears the extra strength around Q=-127 $\sim$ -128 MeV due to the [(2p$_{3/2}$)$_{n}^{-1}\otimes$(1s)$_{\pi}$] and [(2p$_{1/2}$)$_{n}^{-1}\otimes$(1s)$_{\pi}$] contributions, which are negligible for $F$=1.0 case because of the smaller momentum transfer.

We also investigate the case of Xe, which is known to have large formation
cross section for pionic 1s state coupled to the s$_{1/2}$ neutron state
\cite{umemoto00}.
For this case, we used the radius parameters,
$R_0$, $R_{ls}$ and $R_c$, for the proton and neutron potential listed
in Table \ref{table1} by assuming the mass number dependence A$^{1/3}$. 
The other parameters are assumed to be the same.
We show the calculated binding energies in Table \ref{Xe-b.e.}. We do not
have the bound 1h$_{11/2}$ state for the value of $F$ smaller than 0.775. The calculated root mean square radii are shown in Table \ref{Xe-rms}. We
find again that the long halo-like tail is developed well for 3s$_{1/2}$ state, the state with the lowest angular momentum, as in the case of Pb.

\begin{figure}[h]
\begin{center}
\includegraphics[height=8cm]{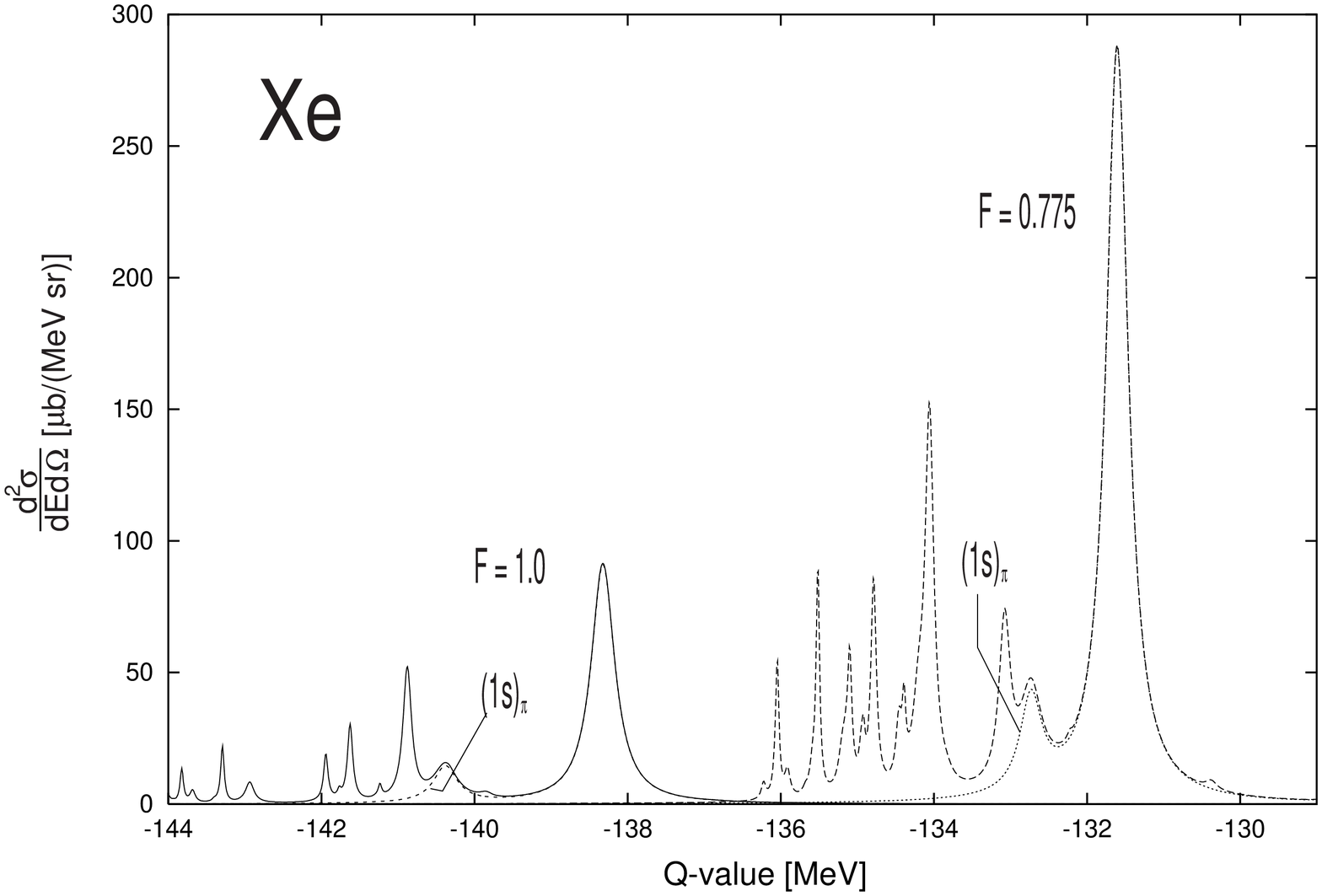}
 \caption{Formation cross sections of the bound pionic states in (d,$^3$He) reaction are plotted as a function of Q-value at the incident energy T$_d$=500 MeV on Xe.
The solid and the dashed curves are the results with $F$=1.0 and 0.775 respectively. The largest peak corresponds to the pionic 1s state, and then we show the separate contributions coming from the pionic 1s components as dotted curves. Instrumental energy resolution is assumed to be 50 keV.
}
\label{fig:Xe-spect}
\end{center}
\end{figure}

We show the calculated spectra for the case of Xe target in Fig. \ref{fig:Xe-spect} as a function of the reaction Q-value. We can see again that there exists large enhancement of the cross section for halo-like neutron states. We also find the change of the spectrum as in the case of Pb. The formation cross sections of [(3s$_{1/2}$)$_{n}^{-1}\otimes$(1s)$_{\pi}$] configuration, which has the largest contribution to the spectra, is around factor 2.9 larger for $F$=0.775 than that for $F$=1.0 which corresponds to no-halo case.

\begin{figure}[h]
\begin{center}
\includegraphics[height=8cm]{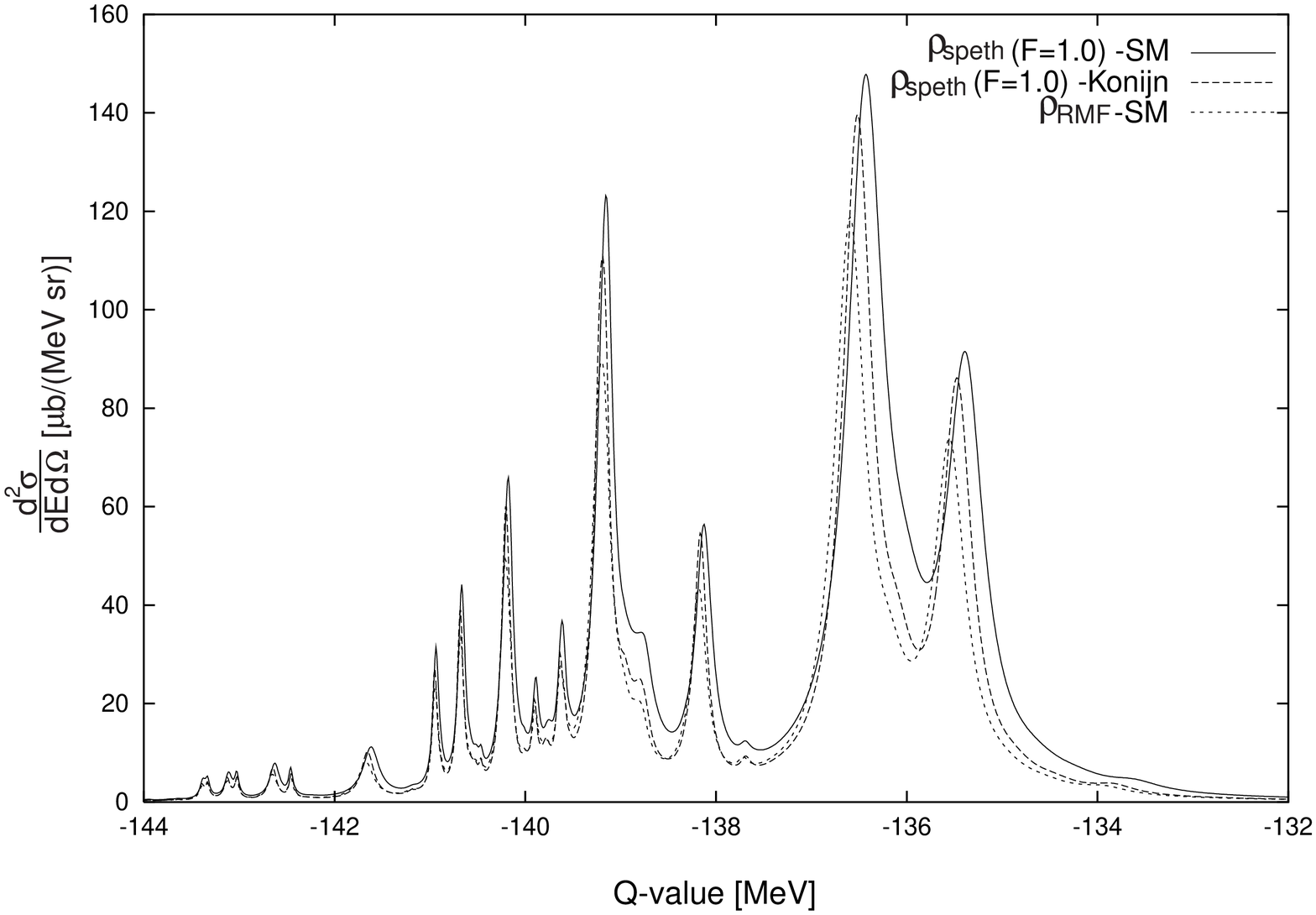}
 \caption{Formation cross sections of the bound pionic states in (d,$^3$He) reaction are plotted as a function of Q-value at the incident energy T$_d$=500 MeV on Pb. Each line shows the calculated spectra with the different combination of the nuclear density and the pion-nucleus optical potential as indicated in the figure. All densities do not have the halo structure. Solid line indicates the same result as the solid line in Fig. \ref{fig:Pb-spect}. Instrumental energy resolution is assumed to be 50 keV. See text for the detail.
}
\label{fig:f=1.0-spect}
\end{center}
\end{figure}

\begin{figure}[h]
\begin{center}
\includegraphics[height=8cm]{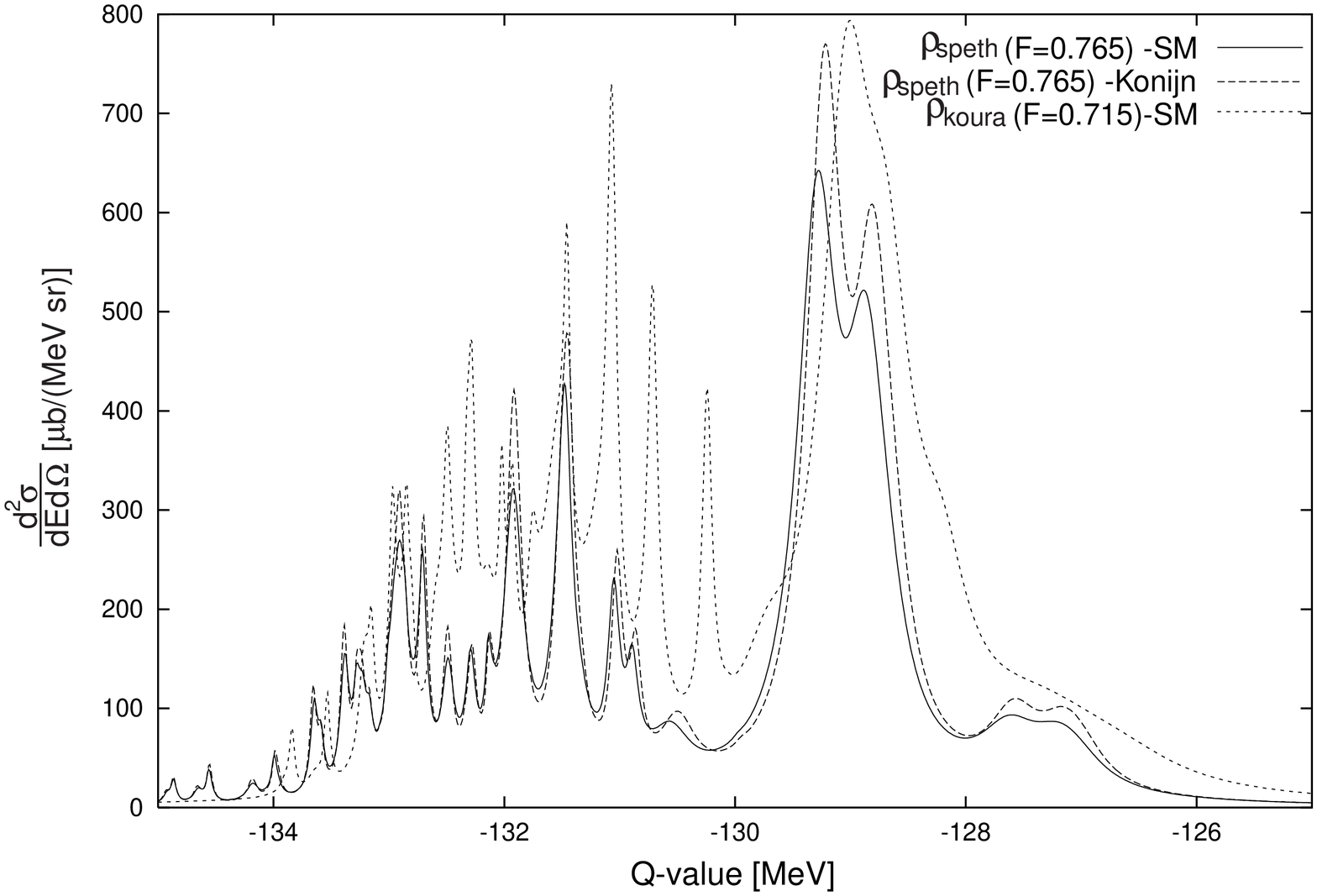}
 \caption{Formation cross sections of the bound pionic states in (d,$^3$He) reaction are plotted as a function of Q-value at the incident energy T$_d$=500 MeV on Pb. Each line shows the calculated spectra with the different combination of the nuclear density and the pion-nucleus optical potential as indicated in the figure. All densities are expected to simulate halo structure.  Solid line indicates the same result as the dashed line in Fig. \ref{fig:Pb-spect}. Instrumental energy resolution is assumed to be 50 keV. See text for the detail.
}
\label{fig:f=0.715-spect}
\end{center}
\end{figure}

Finally, we examined the sensitivity of our results 
to the theoretical inputs. In Fig. \ref{fig:f=1.0-spect}, 
we show the (d,$^3$He) spectra for Pb which are 
calculated by using the nuclear density with the 
relativistic mean field model(RMF) \cite{daisy} and 
with the pion-nucleus optical potential obtained by 
Konijn et al. \cite{konijn}. The RMF model gives 
the nuclear density without the halo-like structure. 
For comparison, we also show the results 
calculated by the nuclear density with the potential 
parameters in  Table 1 denoted as $\rho_{\rm speth}$ \cite{speth77} and/or 
the results with the pion-nucleus optical potential 
by Seki and Masutani \cite{seki83} (SM).  As seen, the calculated 
spectra are slightly modified by the above choice of 
the nuclear densities and the pion-nucleus optical 
potential. But these modifications are not significant 
for the cases considered here.
Next, in Fig. \ref{fig:f=0.715-spect}, we show the 
results by assuming the halo-like nuclear densities.   
We adopted the same procedure as in Section 2 to 
simulate the possible halo-like structure of the 
nuclear density. The strength of the single-nucleon 
potential is varied by introducing a single free
parameter F as in Eq.(\ref{eqn:factor}). The nucleon single-particle 
potential by Koura (set OB) \cite{koura} is adopted to 
see the dependence on the nucleon-nucleus 
potential. For this potential, we found F=0.715 is the 
smallest F value which gives the bound 1i$_{13/2}$ single-particle state. 
Since the overlap of the pion wave function and the 
nuclear density is larger for this case, we obtained significantly larger widths for 
the deeply bound pionic states. Thus, the calculated 
spectrum with the nuclear density by the single 
nucleon potential by Koura is influenced as seen 
in Fig. \ref{fig:f=0.715-spect}: (i) the largest two peaks due to [(2p$_{3/2}$)$_{n}^{-1}\otimes$(2p)$_{\pi}$] and 
[(2p$_{1/2}$)$_{n}^{-1}\otimes$(2p)$_{\pi}$] configurations can not be distinguished, and (ii) the bump at Q=-127 $\sim$ -128MeV due to pionic 1s-states contribution is not seen with the nuclear density by Koura. 
However, we can see that the absolute heights of the peaks 
in the spectra are roughly the same. 
We can conclude that, though the detailed shape of the 
spectrum is influenced by the theoretical inputs, the formation cross 
section of the deeply bound pionic states is largely enhanced by the existence of the halo-like structure in the 
target nucleus.


\begin{table}[h]
\begin{center}
\begin{tabular}{|c|l|l|l|l|l|l|}
\cline{1-7}
      B.E.(keV) & 3p$_{1/2}$ & 2f$_{5/2}$ & 3p$_{3/2}$ & 1i$_{13/2}$ & 2f$_{7/2}$ &
      1h$_{9/2}$ \\
\cline{1-7}
$F$=1.0 & 7761.6 & 8386.0 & 8807.6 & 8895.0 & 11252.3 & 11868.5\\
\cline{1-7}
$F$=0.765 & 1094.4 & 933.91 & 1513.8 & 18.281 & 2787.6 & 2864.4\\
\cline{1-7}
\end{tabular}
\caption{Calculated neutron binding energies for each valence neutron level of Pb in unit of keV for $F$=1.0 and 0.765.}
\label{Pb-b.e.}
\end{center}
\end{table}


\begin{table}[h]
\begin{center}
\begin{tabular}{|c|l|l|l|l|l|l|}
\cline{1-7}
$<r^2>^{1/2}$(fm)& 3p$_{1/2}$ & 2f$_{5/2}$ & 3p$_{3/2}$ & 1i$_{13/2}$ & 2f$_{7/2}$ &
      1h$_{9/2}$ \\
\cline{1-7}
$F$=1.0 & 5.9145 & 5.7722 & 5.7723 & 6.2341 & 5.6795 & 5.6008\\
\cline{1-7}
$F$=0.765 & 7.9384 & 6.9139 & 7.4099 & 6.5740 & 6.3475 & 5.9810\\
\cline{1-7}
\end{tabular}
\caption{Calculated root mean square radii for each valence neutron state of Pb in unit of fm for $F$=1.0 and 0.765.}
\label{Pb-rms}
\end{center}
\end{table}

\begin{table}[h]
\begin{center}
\begin{tabular}{|c|l|l|l|l|l|}
\cline{1-6}
B.E.(keV)& 1h$_{11/2}$ & 3s$_{1/2}$ & 2d$_{3/2}$ & 1g$_{7/2}$ & 2d$_{5/2}$ \\
\cline{1-6}
$F$=1.0 & 8215.9 & 9668.8 & 9219.1 & 11836.0 & 11726.4 \\
\cline{1-6}
$F$=0.775 & 121.0 & 2821.6 & 2306.4 & 3579.1 & 3946.9 \\
\cline{1-6}
\end{tabular}
\caption{Calculated neutron binding energies for each valence neutron level of Xe
in unit of keV for $F$=1.0 and 0.775.}
\label{Xe-b.e.}
\end{center}
\end{table}

\begin{table}[h]
\begin{center}
\begin{tabular}{|c|l|l|l|l|l|}
\cline{1-6}
$<r^2>^{1/2}$(fm)& 1h$_{11/2}$ & 3s$_{1/2}$ & 2d$_{3/2}$ & 1g$_{7/2}$ & 2d$_{5/2}$ \\

\cline{1-6}
$F$=1.0 & 5.4817 & 5.1079 & 5.0575 & 4.8649 & 4.9911 \\
\cline{1-6}
$F$=0.775 & 5.8531 & 6.2414 & 6.0258 & 5.2500 & 5.614 \\
\cline{1-6}
\end{tabular}
\caption{Calculated root mean square radii for each valence neutron state of Xe in unit of fm
for $F$=1.0 and 0.775.}
\label{Xe-rms}
\end{center}
\end{table}

\section{Conclusions}

In this paper, we have studied theoretically the formation cross sections of the deeply bound pionic states in (d,$^3$He) reactions for halo nuclei. To simulate the halo-like structure of neutrons, we have simply modified the depth of the neutron potential by introducing a single free parameter $F$ and generated the shallow neutron states with long tail. Using the neutron wave function and the nuclear densities thus obtained, we have calculated the formation cross section of the deeply bound pionic states in (d,$^3$He) reactions. By varying the parameter $F$, we can see the effect of valence neutron states on the formation spectra. We found that the formation cross section is significantly enhanced for the case with neutron halo-like states. We have examined the several theoretical models and found that the enhancement of the reaction strength due to the halo-like structure in the target nucleus is the common feature of the spectrum. For the cases with halo-like states, the reaction strengths are enhanced about a factor 3.5 for Pb and about 2.9 for Xe than the case without neutron halo. We also found that the shape of the reaction spectra is considerably affected by the choice of the strength parameter $F$ for the nucleon-nucleus potential.

We believe that the present results are relevant to the development of spectroscopic studies for the deeply bound pionic states in wide area in nuclear chart including the neutron-rich unstable nuclei.

\subsection*{Acknowledgments}

One of us (M.F.) thanks Dr.H. Nagahiro for valuable comments.
This work is
partly supported by the Grants-in-Aid for Scientific Research of the Japan
Ministry of
Education, Culture, Sports, Science and Technology (No. 11694082 and No. 14540268).

\end{document}